\pdfoutput=1
\documentclass[11pt]{article}
\usepackage[margin=1in]{geometry}
\usepackage[T1]{fontenc}
\usepackage[utf8]{inputenc}
\usepackage{lmodern}
\usepackage{microtype}
\usepackage{cite}
\usepackage{amsmath,amssymb,amsfonts}
\usepackage{graphicx}
\usepackage{textcomp}
\usepackage{booktabs}
\usepackage{multirow}
\usepackage{array}
\usepackage{url}
\usepackage{bm}
\usepackage[hidelinks]{hyperref}

\hypersetup{
  pdftitle={SwarmReconGuard: Black-Box Detection of Distributed Collective Reconnaissance by Individually Benign-Looking Agent Populations},
  pdfauthor={Vahid Tavakkoli, Kabeh Mohsenzadegan, and Kyandoghere Kyamakya},
  pdfsubject={Distributed collective reconnaissance detection},
  pdfkeywords={agentic AI security, anomaly detection, distributed reconnaissance, multi-agent systems}
}

\begin{document}

\title{SwarmReconGuard: Black-Box Detection of Distributed Collective Reconnaissance by Individually Benign-Looking Agent Populations}

\author{%
	Vahid Tavakkoli$^{1}$,Kabeh Mohsenzadegan $^{1,*}$, Kyandoghere Kyamakya$^{1}$\\[0.6em]
	\small $^{1}$Institute for Smart System Technologies, University of Klagenfurt, Klagenfurt, Austria\\
	\small $^{*}$Corresponding author: Vahid Tavakkoli (\href{mailto:vahid.tavakkoli@aau.at}{vahid.tavakkoli@aau.at})
}
\date{}

\maketitle

\begin{abstract}
Autonomous and agentic clients can distribute reconnaissance across many identities so that each request remains valid, low-rate, and benign-looking while the population collectively acquires broad system knowledge. We formalize this threat as Distributed Collective Reconnaissance (DCR) and present SwarmReconGuard, a reproducible black-box benchmark in which the defender observes only service-boundary telemetry. The Docker-isolated study evaluates 11 benign and attack behaviors across 10–10,000 virtual identities, comprising 440 test runs and 3,666,300 requests, with complete telemetry integrity. We compare semantic, Gaussian, conditional, graph, kernel, hybrid, and CUSUM-based detectors. Gaussian likelihood-ratio detection achieves 100\% detection with 0\% observed false positives on known attacks but only 3\% on unseen policies. CUSUM yields 36.1\% overall detection at 1.25\% false positives, while hybrid CUSUM reaches 85.7\% detection with 0\% observed false positives at 10,000 identities. Results expose a major policy-generalization gap and motivate exposure-aware, scale-aware defenses.
\end{abstract}

\noindent\textbf{Keywords:} agentic AI security; anomaly detection; collective behavior; CUSUM; distributed reconnaissance; intrusion detection; multi-agent systems; semantic exposure; swarm security.

\section{Introduction}
\label{sec:introduction}
\textbf{Modern} public services increasingly expose structured application programming interfaces (APIs) to browsers, mobile applications, automation, and software agents. Conventional defenses are effective when harmful behavior is locally conspicuous: a client exceeds a request-rate limit, an exploit signature appears in payloads, authentication fails repeatedly, or resource consumption resembles denial of service. A more difficult regime arises when a large population performs individually acceptable actions that compose into harmful collective knowledge acquisition. A single request may disclose almost nothing; thousands of strategically distributed requests may reconstruct endpoint behavior, identifier structure, diagnostic metadata, or authorization boundaries.

The emergence of agentic artificial intelligence increases the practical relevance of this problem. Recent work documents the rapid development of autonomous and multi-agent systems, their use in cyber threat intelligence, automated vulnerability analysis, incident response, and security operations, and the accompanying security risks \cite{chhabra2026agentic,addis2026llmagents,moralles2026agentic,shah2025madcti,hmimou2025multiagent,toprani2025iac,bowers2025codeinjection}. At the same time, intrusion-detection research remains dominated by flow-, host-, or identity-level classification, DDoS detection, botnet signatures, learned traffic models, and federated variants \cite{jayalaxmi2022iot,suryotrisongko2022dga,kalakoti2022botnet,mahmoodi2023federated,shurbaji2025botnet,ennaji2025adversarial}. These approaches are valuable but do not directly answer the following question:

\begin{quote}
Can a high-availability service detect coordinated information acquisition when every individual identity remains within ordinary request budgets and the defender cannot observe the agents' internal coordination?
\end{quote}

We call this threat class \emph{Distributed Collective Reconnaissance} (DCR). DCR differs from classical DDoS in objective and observability. The goal is not necessarily to reduce service availability but to maximize cumulative information acquired about the service while keeping marginal behavior plausible. It also differs from a traditional scan performed by one source because attribution can be distributed over many identities, and from a white-box multi-agent monitoring problem because the defender sees only boundary events.

This paper introduces \textbf{SwarmReconGuard}, a reproducible Docker-isolated experimental framework for measuring this regime. The framework is intentionally black-box from the defender's perspective: it observes request/response telemetry at an API gateway but not agent prompts, shared memory, coordination messages, model activations, true swarm membership, or scenario labels. The study uses scripted virtual identities rather than claiming that all 10,000 entities are large-language-model (LLM) agents. Scripted policies provide experimental control at population scale; the threat model is designed to include autonomous agents as a motivating and future deployment class.

The paper makes five main contributions.

\begin{enumerate}
\item We formalize DCR as a collective information-acquisition problem under per-identity validity, hidden coordination, and service-availability constraints, and state an observability limit that prevents any boundary-only detector from recovering hidden intent when benign and adversarial observable distributions are identical.
\item We present a reproducible benchmark with four benign controls, seven attack policies, four population scales, independent training/calibration/test phases, exact telemetry accounting, raw-event replay, and an internal-only safety boundary.
\item We evaluate ten online detector families and three ablations, ranging from transparent semantic features to distributional, graph, kernel, fusion, and sequential evidence-accumulation methods.
\item We evaluate operational protection with false-positive rate (FPR), detection, precision, discrimination, delay, and \emph{Exposure Before Detection} (EBD), while counting missed attacks through exposure-at-alarm-or-end rather than excluding them from exposure summaries.
\item We provide a generalization study across known versus unseen attack policies and population scales. The strongest known-policy result is near-perfect, yet unseen-policy recall falls sharply. This negative result is central: excellent known-attack separation does not establish robust DCR defense.
\end{enumerate}

The remainder of the paper reviews related work, formalizes the threat model, describes SwarmReconGuard and its detectors, reports the experimental protocol and results, and discusses implications, limitations, and research directions.

\section{Related Work and Research Gap}
\label{sec:related}

\subsection{Network Intrusion, Botnet, and DDoS Detection}
Recent IEEE literature has substantially advanced machine-learning-based intrusion detection. Surveys have reviewed ML/DL approaches for IoT intrusion prevention \cite{jayalaxmi2022iot}, IoT botnet detection \cite{shurbaji2025botnet}, Internet-of-Drones IDS \cite{ogab2025drones}, and adversarial robustness of network IDS \cite{ennaji2025adversarial}. Representative systems use statistical feature selection for botnets \cite{kalakoti2022botnet}, source-side collaborative sequence models for DDoS \cite{yeom2022ddos}, network-traffic visualization \cite{arnold2024traffic}, explainability frameworks \cite{arreche2024exai,gaspar2024lime}, and deep or federated learning \cite{mahmoodi2023federated,venkatasubramanian2023federated,baich2025federated,gutti2025federated,hamad2025systematic}. SDN-focused work also includes targeted control-plane defenses \cite{liu2023poaguard} and privacy-preserving intrusion detection frameworks \cite{kamaruddin2026felv}. DDoS-specific work emphasizes volumetric or protocol-level anomalies and service degradation \cite{hirsi2025ddos,taniguchi2026cdn}. These methods provide important baselines, but DCR may preserve normal aggregate throughput and deliberately distribute semantic exploration across clients.

Dataset realism and reproducibility are also recurring concerns. Edge-IIoTset introduced a purpose-built IoT/IIoT cybersecurity testbed \cite{ferrag2022edge}, while Arifin \emph{et al.} developed a SCADA dataset generated from a hybrid physical/virtual environment \cite{arifin2024scada}. SwarmReconGuard follows the same principle that controlled, replayable testbeds are necessary, but targets a different dependent variable: collective semantic coverage and information exposure rather than only packet/flow attack class.

\subsection{Zero Trust, Context, and Boundary Enforcement}
Zero-trust architectures emphasize continuous verification, least privilege, and resource-centered policy rather than implicit trust \cite{syed2022zerotrust}. Later work has explored zero-trust hardware/supply-chain enforcement \cite{kulkarni2024zerotrust} and session-lifecycle-aware authorization with rapid revocation \cite{chang2026session}. These ideas are complementary to DCR defense because contextual authorization can add differentiating information when pure behavioral observation is insufficient. SwarmReconGuard does not implement authorization mitigation; instead it measures what can be inferred from traffic when valid requests have already been admitted.

\subsection{Distribution Shift, Graphs, and Explainability}
Network behavior is rarely stationary or exactly Gaussian. Dynamic graph representation learning provides a general framework for evolving relational structures \cite{yang2024dynamic}; log anomaly detection illustrates sequence-level modeling in operational data \cite{aziz2024logs}; and explainable IDS research emphasizes the need to understand model decisions and failure modes \cite{arreche2024exai,gaspar2024lime}. Our detector suite therefore includes both an interpretable heuristic and deliberately simple statistical baselines, while treating Gaussianity as an approximation to be tested rather than assumed. We also include handcrafted temporal bipartite graph descriptors and a kernel distributional baseline to examine whether cross-identity organization is detectable without reconstructing a hidden swarm graph.

\subsection{Agentic and Multi-Agent Cybersecurity}
Agentic AI now supports cyber threat intelligence, vulnerability assessment, incident response, scenario generation, and security operations. MAD-CTI uses LLM agents for dark-web threat analysis \cite{shah2025madcti}; Hmimou \emph{et al.} use multiple agents for threat detection and correlation \cite{hmimou2025multiagent}; Toprani and Madisetti apply agentic workflows to infrastructure-as-code vulnerabilities \cite{toprani2025iac}; Roh and Kim generate cyberattack scenarios with LLMs \cite{roh2025scenario}; AutoBnB studies multi-agent incident response \cite{liu2025autobnb}; Cognitive SOC coordinates LLM agents in security operations \cite{sheikhi2025soc}; and recent studies expose code-injection risks in LLM-based multi-agent software development \cite{bowers2025codeinjection}. Broad surveys now characterize agentic-AI architectures and security risks \cite{moralles2026agentic,chhabra2026agentic,addis2026llmagents}.

Recent multi-agent safety research is closer to our collective-risk setting, but differs in observability. NARCBench studies covert collusion with white-box activation probes and reports a marked transfer gap under structurally different scenarios \cite{rose2026narcbench}. RiskLab provides a controlled toolkit for emergent risks such as collusion and resource overreach in LLM multi-agent systems \cite{jiang2026risklab}, while long-horizon experiments show that undesirable coordination can emerge from repeated agent interaction \cite{shi2026collusion}. ALTEDA detects and attributes multi-agent LLM threats from synchronized application, network, and system logs without reading prompts or responses \cite{rabieinejad2026alteda}. These studies establish collective and interaction-level risk as a serious problem. SwarmReconGuard complements them by assuming that the attacker-side runtime, inter-agent messages, and internal logs are inaccessible: the defender receives only the service-boundary footprint of the population.

This literature shows both the utility and the attack potential of autonomous agents, but most work observes the agent framework itself or evaluates task performance. SwarmReconGuard instead assumes the attacker-side runtime is inaccessible. The research gap is therefore the \emph{boundary-only} detection of a harmful collective whose members can each remain policy-compliant.

\subsection{Research Gap}
The closest conventional analogies are low-and-slow scanning, distributed botnets, API enumeration, DDoS, and anomaly detection. None alone captures all four properties that define our target regime: (i) individually valid low-budget actions, (ii) hidden coordination or complementary exploration, (iii) cumulative semantic coverage rather than availability loss as the attacker objective, and (iv) a service-side defender limited to boundary telemetry. SwarmReconGuard operationalizes this combination and explicitly measures policy generalization rather than only in-distribution accuracy.

\section{Threat Model and Problem Formulation}
\label{sec:threat}

\subsection{Service, Agents, and Observables}
Let a service have hidden state $S$ and an exposed request mechanism $M$. At step $t$, an identity $i_t$ issues an allowed query $q_t$ and receives response
\begin{equation}
y_t=M(S,q_t).
\end{equation}
The attacker may control a set of identities $\mathcal{A}$ and coordinate them outside the defender's environment. The defender does not observe hidden policy state, prompts, shared memory, internal messages, or the true membership of $\mathcal{A}$. It observes only boundary telemetry
\begin{equation}
o_t=\phi(q_t,y_t,m_t),
\end{equation}
where $m_t$ denotes timing, synthetic identity, status, body size, endpoint family, and related service metadata.

DCR is an attack process in which the population maximizes cumulative information acquisition while each member remains locally plausible. An information-theoretic idealization is
\begin{equation}
\max_{\pi_A} I(S;Y_{1:T})
\quad \text{s.t.} \quad
R_i(A_i)\leq \tau_i, \; \mathrm{SLA}\geq \mathrm{SLA}_{\min},
\label{eq:attacker}
\end{equation}
where $R_i$ is an individual risk/rate rule. In the synthetic benchmark, exact mutual information is not available; we use measurable semantic coverage and synthetic weakness discovery as proxies.

Let $\Phi(q)$ be the semantic resource key revealed by a query and $\mathcal{S}_{obs}$ the finite synthetic observable namespace. Collective semantic coverage is
\begin{equation}
C(t)=\frac{|\bigcup_{j\leq t}\Phi(q_j)|}{|\mathcal{S}_{obs}|}.
\label{eq:coverage}
\end{equation}
The operational defender seeks an alarm rule $D$ that reduces exposure while constraining false alarms and service cost:
\begin{equation}
\min_D \; \mathbb{E}[E_D]+\alpha\,\mathrm{FPR}+\beta\,\Delta\mathrm{SLA},
\label{eq:defender}
\end{equation}
where $E_D$ is exposure at alarm, or final exposure when an attack is missed.

\subsection{Observability Limit}
A boundary-only detector cannot infer hidden coordination if the attacker exactly matches the benign observable distribution. Let $P_B(O)$ and $P_A(O)$ be distributions over complete observable streams. With equal priors, the optimal classification accuracy is
\begin{equation}
\mathrm{Acc}^{*}=\frac{1}{2}\left(1+\mathrm{TV}(P_A,P_B)\right),
\end{equation}
where $\mathrm{TV}$ is total variation distance. If $P_A=P_B$, then $\mathrm{TV}=0$ and no detector restricted to $O$ can outperform chance. This elementary decision-theoretic limit is important: successful DCR defense must exploit an observable distributional difference, introduce new differentiating context (e.g., stronger authorization or proof-of-context), reduce disclosure, or accept residual leakage. SwarmReconGuard evaluates the first option.

\subsection{Attacker and Defender Assumptions}
The benchmark attacker can distribute valid HTTP GET requests among virtual identities and, depending on scenario, coordinate complementary namespace exploration. It cannot access host shells, real secrets, external systems, or the detector. The defender owns the gateway and target service and sees complete synthetic boundary telemetry. It does not receive scenario labels at inference time, and mixed-swarm membership is evaluator-only. Trial boundaries are supplied by the laboratory; continuous production-stream segmentation remains outside this work.

\section{SwarmReconGuard Architecture}
\label{sec:architecture}

\begin{figure}[t]
\centering
\includegraphics[width=0.94\textwidth]{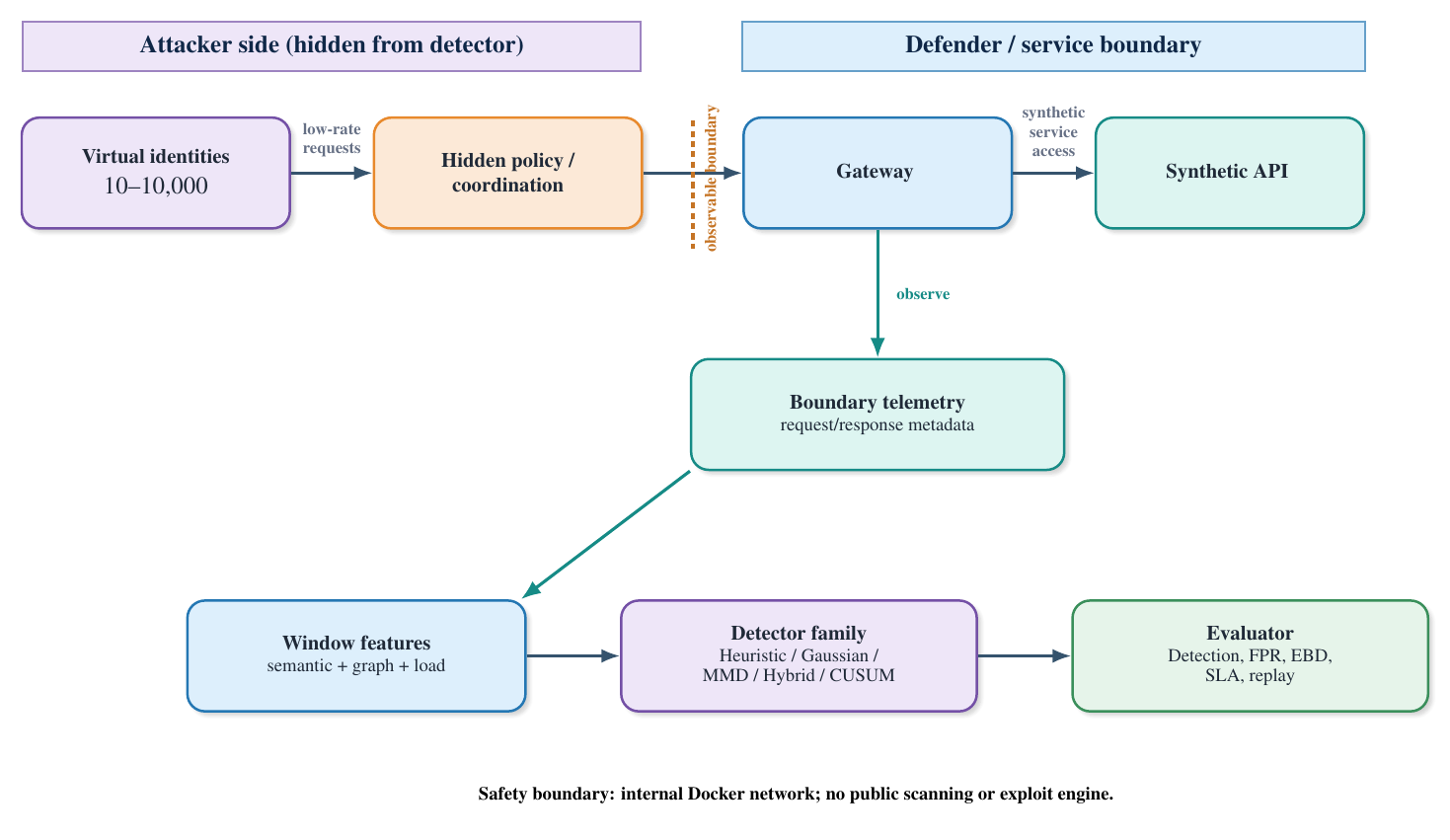}
\caption{SwarmReconGuard architecture. Attacker-side policy and coordination are hidden. The defender receives only API-boundary telemetry, from which window-level semantic, graph, distributional, and sequential evidence is derived.}
\label{fig:architecture}
\end{figure}

Figure~\ref{fig:architecture} summarizes the design. The environment comprises an internal Docker network, a synthetic municipal-style API, a gateway that records telemetry, a Redis event stream, online detectors, an experiment runner, and an evaluator/reporting pipeline. No public target port is exposed, the experiment runner rejects arbitrary non-lab hosts, Linux capabilities are dropped, no Docker socket is mounted, and the target contains no shell-execution or real-exploitation path.

\subsection{Synthetic Information Channels}
The service contains three harmless information surfaces selected because they admit measurable ground truth without real exploitation:
\begin{itemize}
\item \textbf{V001 -- ticket enumeration:} valid and invalid synthetic ticket identifiers reveal existence/status structure;
\item \textbf{V002 -- permit-prefix oracle:} a two-character prefix query returns a synthetic aggregate count over a finite namespace;
\item \textbf{V003 -- diagnostic disclosure:} a valid diagnostic view returns unnecessary synthetic backend metadata.
\end{itemize}
The system records discovery of these conditions and population/attacker semantic coverage. Coverage is explicitly interpreted as a proxy for information acquisition, not as a percentage of real secrets reconstructed.

\subsection{Traffic Regimes}
Table~\ref{tab:scenarios} lists the 11 evaluation scenarios. Four benign workloads are designed to prevent convenient shortcuts: legitimate diagnostic access removes an exclusive diagnostic cue; benign exploration creates broad semantic novelty; and benign bulk access creates a high-volume hard negative. Seven attack policies include two known training policies and five unseen variants. The matched-swarm control approximately aligns individual resource marginals while changing collective sampling structure. The mixed-swarm scenario embeds a hidden 20\% attacker subset inside benign traffic.

\begin{table}[t]
\centering
\caption{Benchmark Scenarios}
\label{tab:scenarios}
\scriptsize
\begin{tabular}{>{\raggedright\arraybackslash}p{0.22\textwidth}>{\raggedright\arraybackslash}p{0.10\textwidth}>{\raggedright\arraybackslash}p{0.58\textwidth}}
\toprule
\textbf{Scenario} & \textbf{Label} & \textbf{Purpose} \\
\midrule
benign\_flash & Benign & Locality, popular resources, and repeated ordinary access. \\
benign\_diagnostic & Benign & Legitimate diagnostic access; removes an attack-exclusive metadata cue. \\
benign\_explorer & Benign & Independent broad exploration with low diagnostic probability. \\
benign\_bulk & Benign & Broad repetitive batch workflow and high-volume hard negative. \\
independent\_recon & Known attack & Independent uniform reconnaissance; one of two attack policies used for fitting. \\
coordinated\_swarm & Known attack & Randomized stratified partitioning that reduces duplicate work. \\
swarm\_no\_diagnostic & Unseen attack & Coordinated exploration without the diagnostic channel. \\
matched\_swarm & Unseen attack & Approximate per-identity marginal matching with coordinated collective sampling. \\
camouflaged\_swarm & Unseen attack & Mixes hot-resource requests with coordinated exploration. \\
low\_and\_slow\_swarm & Unseen attack & Adds seeded inter-request delay to reduce temporal conspicuity. \\
mixed\_swarm & Unseen attack & Hides a seeded 20\% attacker subset inside benign population traffic. \\
\bottomrule
\end{tabular}
\end{table}

\section{Detection Methods}
\label{sec:methods}

All online methods score identical non-overlapping event windows. The first window contains 16 events where available; subsequent publication windows contain up to 64 events with a one-second time limit, and the terminal tail is explicitly scored so events are never silently discarded.

\subsection{Transparent Semantic Heuristic}
For a window, SwarmReconGuard computes novelty ratio $N$, normalized namespace span $S$, gap uniformity $G$, diagnostic ratio $D$, endpoint-family entropy $H$, and semantic coverage pressure $C$. The transparent baseline is
\begin{equation}
R_h=0.18N+0.22S+0.20G+0.15D+0.15H+0.10C.
\label{eq:heuristic}
\end{equation}
The purpose is not to claim optimal weights but to provide an auditable baseline that embodies the hypothesis that systematic exploration has semantic structure beyond request volume.

\subsection{Gaussian and Conditional Distribution Models}
Bounded features are clipped and logit transformed. The unconditional benign model is a shrinkage-regularized multivariate Gaussian
\begin{equation}
\bm{x}\sim\mathcal{N}(\bm{\mu}_B,\bm{\Sigma}_B),
\end{equation}
with squared Mahalanobis OOD score
\begin{equation}
d_B^2(\bm{x})=(\bm{x}-\bm{\mu}_B)^T\bm{\Sigma}_B^{-1}(\bm{x}-\bm{\mu}_B).
\end{equation}
Known attack policies are represented as a mixture, producing a likelihood-ratio score
\begin{equation}
\Lambda(\bm{x})=\log p_A(\bm{x})-\log p_B(\bm{x}).
\label{eq:llr}
\end{equation}

Because benign behavior changes with workload and scale, the conditional models regress transformed features on context
\begin{equation}
\bm{c}=[1,\log(1+n_w),\log(1+i_w),\log(1+n_c),\log(1+i_c)],
\end{equation}
where $n_w/i_w$ and $n_c/i_c$ are window/cumulative request and identity counts. Gaussian residual models are then fitted per training regime. This yields conditional OOD and conditional LLR scores.

\subsection{Collective Graph Descriptors}
A temporal bipartite identity--resource-region graph is built over each window. Descriptors include cross-identity novelty, region balance and entropy, identity specialization, temporal namespace progression, connected-component fraction, and cumulative variants. The graph detector is deliberately a statistical model over interpretable descriptors rather than a graph neural network. It tests whether organization such as complementary exploration is observable without access to a true attacker coordination graph.

\subsection{Kernel Distribution Shift}
To relax Gaussian assumptions, benign conditional residuals are embedded with random Fourier features approximating a Gaussian kernel. The detector compares the recent residual-feature mean with a benign reference mean, yielding an approximate kernel maximum mean discrepancy (MMD) score. This is a computational distribution-shift signal rather than an exact permutation-test $p$-value; the conceptual connection is to kernel two-sample testing \cite{gretton2012kernel}.

\subsection{Hybrid and Sequential Evidence}
The pointwise hybrid standardizes and averages three training-derived signals:
\begin{equation}
R_{hyb}=\frac{z(R_h)+z(\Lambda_c)+z(R_g)}{3},
\end{equation}
where $\Lambda_c$ is conditional LLR and $R_g$ is graph OOD evidence. Three ablations remove graph, LLR, or heuristic evidence.

Sequential methods accumulate weak evidence. The LLR CUSUM is
\begin{equation}
C_t=\max(0,C_{t-1}+\Lambda_{c,t}),
\label{eq:cusum}
\end{equation}
and hybrid CUSUM uses
\begin{equation}
C_t^{hyb}=\max(0,C_{t-1}^{hyb}+R_{hyb,t}-\kappa),
\end{equation}
with drift allowance $\kappa=0.5$. CUSUM follows the classical principle of sequential change accumulation \cite{page1954cusum} but is applied here to collective semantic evidence.

\subsection{Independent Calibration}
Training and alarm calibration are separate. Model parameters are fitted on 72 measured HTTP training streams. Thresholds are then calibrated using 299 independent benign HTTP streams. For method $m$, let $M_j^{(m)}$ be the maximum score over calibration stream $j$. With target $\alpha=0.01$, the rank threshold is
\begin{equation}
\tau_m=M^{(m)}_{(k)},\qquad
k=\left\lceil(n+1)(1-\alpha)\right\rceil.
\end{equation}
For $n=299$, $k=297$ and the empirical rank resolution is $1/(n+1)=0.00333$. This calibration controls one alarm decision per complete stream under exchangeability and comparable horizons; it is not a claim of a universal 1\% production FPR at every population scale.

\section{Experimental Design}
\label{sec:experimental}

\subsection{Study Matrix and Integrity}
The test matrix contains 11 scenarios, four population sizes $N\in\{10,100,1000,10000\}$, and 10 repetitions, yielding 440 test runs. Every identity receives three requests. The study therefore generates 3,666,300 test requests. Training and calibration are independent and add 3,066,990 measured reference HTTP requests. The experiment was executed in a Linux/WSL2 environment with 16 logical CPUs and Python 3.12.12; all containers run on the internal Docker network.

Strict telemetry is a design invariant. A run is valid only when the client logical-request count, gateway forwarded count, and detector-processed event count agree and no telemetry failure or unrecovered transport error occurs. The final study contains 440/440 valid runs and exactly 3,666,300 processed events for 3,666,300 test requests, with zero telemetry failures. Bounded transport retry and idempotent logical request identifiers protect measurement integrity without accepting incomplete traces.

\subsection{Metrics}
We report AUROC and average precision (AP) as threshold-independent ranking measures, and detection rate, FPR, precision, F1, and Wilson 95\% intervals at the independently calibrated operating point. Repeated continuous metrics use sample standard deviation and Student-$t$ intervals. Hedges' $g$ is reported for selected effect-size contrasts.

Exposure Before Detection is
\begin{equation}
\mathrm{EBD}=C(t_{alarm}).
\end{equation}
However, averaging EBD only over detected attacks is selection-biased. We therefore also report \emph{exposure at alarm or end}, defined as $C(t_{alarm})$ if detected and $C(T)$ otherwise. Event-prefix alarm exposure is distinguished from wall-clock exposure audited at the actual observer alarm time; backlog can make the latter larger.

Service metrics include client p50/p95/p99 latency, throughput, transport errors, matched benign SLA deltas, detector processing time per event, and processing lag. The detector is observational and does not block requests, so these measurements quantify benchmark overhead and traffic effects rather than mitigation benefit.

\subsection{Offline Generalization Analysis}
Completed-run Gaussian baselines are also evaluated with leave-repeat-out, leave-scale-out, and leave-attack-policy-out folds. An inner benign repetition is excluded from model fitting and used only for threshold calibration. Because independently fitted folds need not share a common score scale, macro-fold AUROC/AP are primary. These smaller folds are supportive analyses; they do not establish the online 1\% FPR target.

\section{Results}
\label{sec:results}

\subsection{Overall Online Operating Points}
Table~\ref{tab:overall} summarizes selected online methods. The central result is not a single universally dominant detector. Different methods occupy different precision--recall--exposure trade-offs. The transparent heuristic achieves AUROC 0.828 with 32.1\% detection at 0.625\% FPR. Gaussian LLR produces zero observed benign alarms and 30.7\% attack detection. CUSUM gives the highest overall detection rate (36.1\%) and F1 (0.527), at 1.25\% FPR and 98.1\% precision. Hybrid CUSUM is more conservative overall but becomes much stronger at the largest population scale.

\begin{table}[t]
\centering
\caption{Selected Online Results Across 440 Test Runs}
\label{tab:overall}
\scriptsize
\setlength{\tabcolsep}{2.5pt}
\resizebox{\textwidth}{!}{%
\begin{tabular}{lrrrrrrrr}
\toprule
\textbf{Method} & \textbf{AUROC} & \textbf{AP} & \textbf{FPR} & \textbf{Detection} & \textbf{Precision} & \textbf{F1} & \textbf{Exposure alarm/end} & \textbf{Median delay} \\
\midrule
Heuristic & 0.828 & 0.895 & 0.625\% & 32.1\% & 98.9\% & 0.485 & 6.15\% & 0.90 s \\
Gaussian LLR & 0.820 & 0.890 & 0.000\% & 30.7\% & 100\% & 0.470 & 6.15\% & 6.45 s \\
Conditional LLR & 0.789 & 0.867 & 0.625\% & 21.1\% & 98.3\% & 0.347 & 6.04\% & 0.85 s \\
CUSUM & 0.787 & 0.873 & 1.250\% & \textbf{36.1\%} & 98.1\% & \textbf{0.527} & \textbf{4.89\%} & 1.11 s \\
Hybrid CUSUM & 0.752 & 0.855 & 0.625\% & 29.3\% & 98.8\% & 0.452 & 5.69\% & 35.74 s \\
Hybrid without graph & \textbf{0.842} & \textbf{0.901} & 0.625\% & 28.9\% & 98.8\% & 0.448 & 6.16\% & \textbf{0.70 s} \\
\bottomrule
\end{tabular}%
}
\end{table}

\begin{figure}[t]
\centering
\includegraphics[width=\columnwidth]{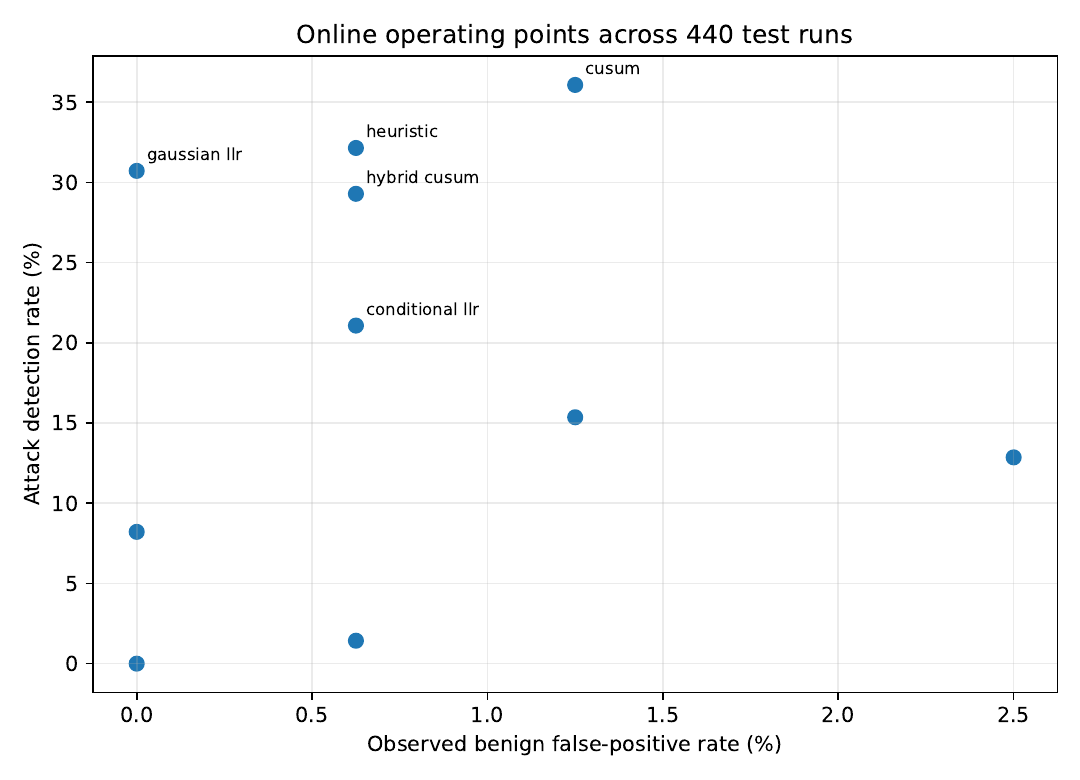}
\caption{Observed operating-point trade-off between false positives and attack detection. Threshold-independent discrimination and calibrated recall are distinct properties.}
\label{fig:tradeoff}
\end{figure}

Figure~\ref{fig:tradeoff} emphasizes that high AUROC does not automatically translate to high recall at a conservative calibrated threshold. This distinction is central for safety-critical deployment: discrimination, calibration, and early protection must be reported separately.

\subsection{Known Versus Unseen Attack Policies}
\begin{figure}[t]
\centering
\includegraphics[width=\columnwidth]{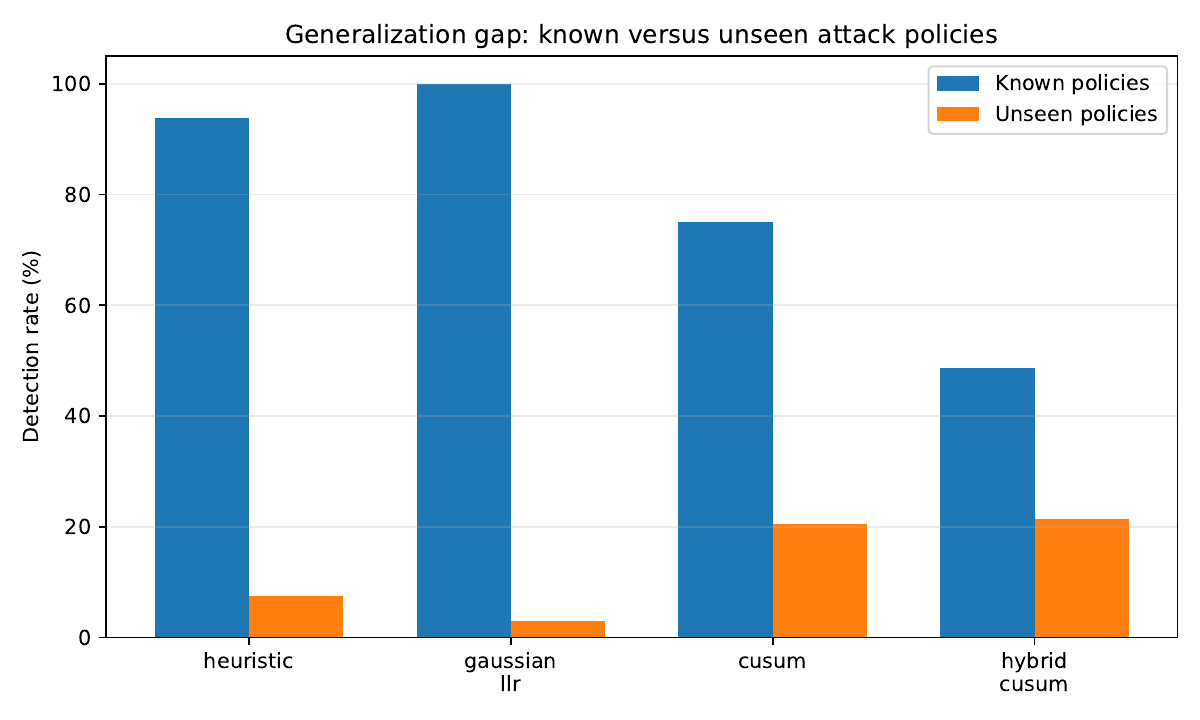}
\caption{Detection rate on known and unseen attack policies. Near-perfect known-policy performance does not imply robustness to policy shift.}
\label{fig:knownunseen}
\end{figure}

The clearest result is the generalization gap shown in Table~\ref{tab:generalization} and Fig.~\ref{fig:knownunseen}. Gaussian LLR perfectly detects the two known attack policies with no observed benign false positives, yet detects only 3.0\% of the five unseen policies. The heuristic falls from 93.75\% known-policy detection to 7.5\% unseen detection. Sequential accumulation partially recovers this gap: CUSUM detects 75.0\% known and 20.5\% unseen attacks, while hybrid CUSUM detects 48.75\% known and 21.5\% unseen attacks. The unseen-policy precision of hybrid CUSUM remains 97.7\%, indicating that the difficulty is recall rather than alarm purity.

\begin{table}[t]
\centering
\caption{Known-Policy Versus Unseen-Policy Generalization}
\label{tab:generalization}
\begin{tabular}{lrrrrrr}
\toprule
\multirow{2}{*}{\textbf{Method}} & \multicolumn{3}{c}{\textbf{Known policies (80 attacks)}} & \multicolumn{3}{c}{\textbf{Unseen policies (200 attacks)}} \\
\cmidrule(lr){2-4}\cmidrule(lr){5-7}
& \textbf{Detection} & \textbf{FPR} & \textbf{AUROC} & \textbf{Detection} & \textbf{FPR} & \textbf{AUROC} \\
\midrule
Heuristic & 93.75\% & 0.625\% & 0.996 & 7.50\% & 0.625\% & 0.761 \\
Gaussian LLR & \textbf{100\%} & \textbf{0\%} & \textbf{1.000} & 3.00\% & \textbf{0\%} & 0.748 \\
CUSUM & 75.00\% & 1.250\% & 0.996 & 20.50\% & 1.250\% & 0.703 \\
Hybrid CUSUM & 48.75\% & 0.625\% & 0.813 & \textbf{21.50\%} & 0.625\% & 0.728 \\
Hybrid without graph & 86.25\% & 0.625\% & 0.998 & 6.00\% & 0.625\% & \textbf{0.780} \\
\bottomrule
\end{tabular}
\end{table}

This result argues against reporting only in-distribution accuracy. DCR strategies can alter coordination, diagnostic usage, timing, or attacker fraction while preserving the high-level objective of collective information acquisition.

\subsection{Population Scale}
\begin{figure}[t]
\centering
\includegraphics[width=\columnwidth]{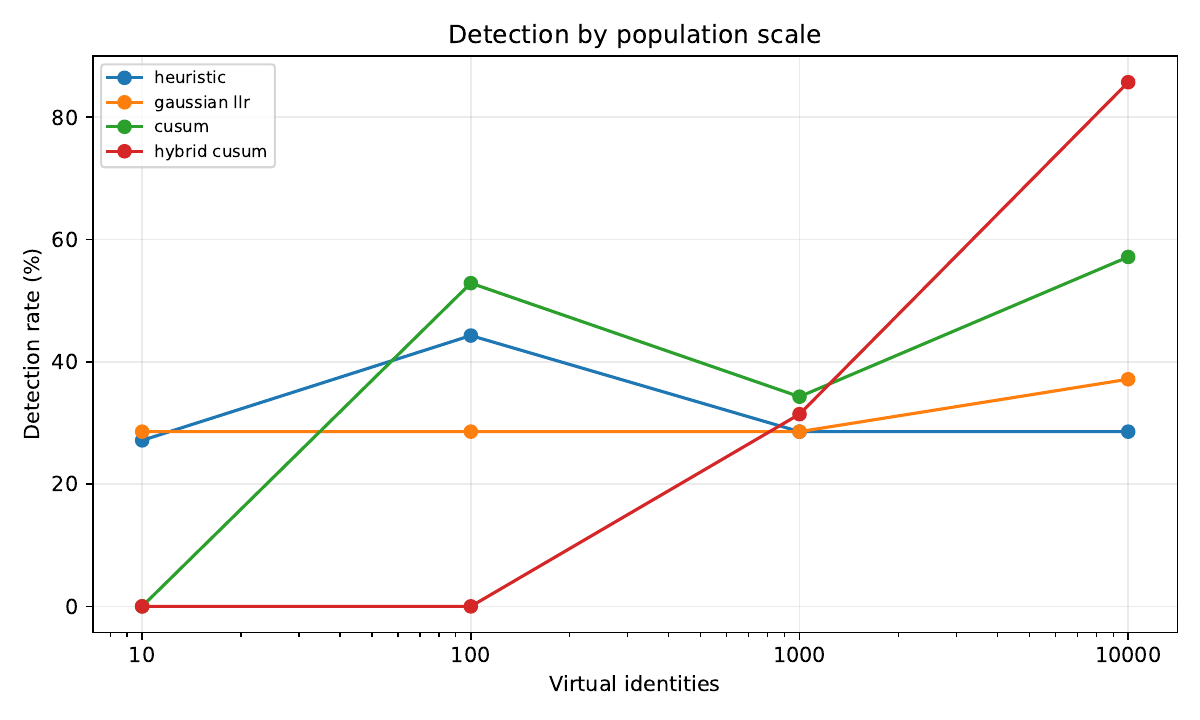}
\caption{Detection rate by population scale for selected methods. The response to scale is detector-dependent; hybrid CUSUM becomes especially effective at 10,000 identities.}
\label{fig:scale}
\end{figure}

Figure~\ref{fig:scale} shows that scale affects methods differently. Hybrid CUSUM is ineffective at 10 and 100 identities, reaches 31.4\% detection at 1,000, and 85.7\% at 10,000 with zero observed benign false positives at that scale. CUSUM detects 52.9\% at 100, 34.3\% at 1,000, and 57.1\% at 10,000. The pointwise heuristic performs best at 100 identities but does not improve monotonically. Thus, ``more agents means easier detection'' is not a universal law; detectability depends on the interaction between policy, windowing, cumulative structure, and calibration. Nevertheless, the largest population produces strong sequential collective evidence.

\subsection{Ablation: Graph Evidence Is Not Yet Helpful}
Ablation results are intentionally reported even though they weaken the proposed fusion. Removing graph evidence increases pointwise hybrid AUROC from 0.674 to 0.842 and detection from 8.2\% to 28.9\%, whereas removing heuristic or LLR evidence collapses performance (Table~\ref{tab:ablation}). This shows that the present handcrafted graph representation is not sufficiently transferable under the hard benign controls and should not be presented as a solved contribution. The result is useful: collective graph structure is theoretically attractive, but the chosen descriptors require refinement before they improve operational detection.

\begin{table}[t]
\centering
\caption{Pointwise Hybrid Ablation}
\label{tab:ablation}
\begin{tabular}{lrrr}
\toprule
\textbf{Variant} & \textbf{AUROC} & \textbf{Detection} & \textbf{FPR} \\
\midrule
Full hybrid & 0.674 & 8.2\% & 0.0\% \\
No graph & \textbf{0.842} & \textbf{28.9\%} & 0.625\% \\
No heuristic & 0.603 & 1.4\% & 0.0\% \\
No LLR & 0.660 & 7.9\% & 0.0\% \\
\bottomrule
\end{tabular}
\end{table}

\begin{figure}[t]
\centering
\includegraphics[width=\columnwidth]{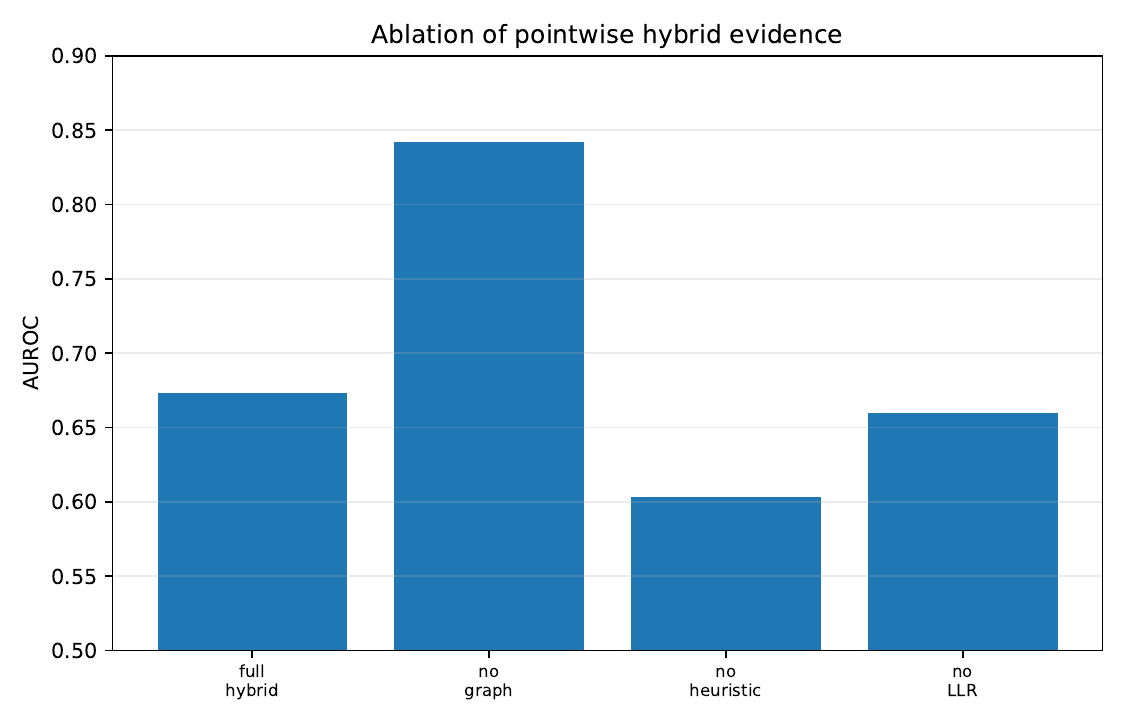}
\caption{AUROC for the pointwise hybrid and component-removal ablations. Removing graph evidence improves performance, identifying a concrete limitation of the current graph descriptor set.}
\label{fig:ablation}
\end{figure}

\subsection{Offline Leave-Attack-Out Validation}
The completed-run offline analysis independently supports the presence of policy shift. In leave-attack-out evaluation, hybrid ranking reaches pooled AUROC 0.841 and macro AUROC 0.846, but its calibrated detection is 39.3\% at 4.46\% FPR. Gaussian LLR yields AUROC 0.809 with 19.6\% detection at 1.79\% FPR. These values are not directly comparable to the stronger 299-stream online calibration because the offline folds have far fewer calibration observations. They reinforce the main conclusion: ranking remains informative under unseen policy, but robust low-FPR operating-point calibration is difficult.

\subsection{Gaussian Assumption Diagnostics}
The benign pooled feature distribution is decisively non-Gaussian. After logit transformation, Shapiro--Wilk tests reject univariate normality for every non-constant feature ($p<10^{-8}$ for semantic coverage and much smaller for several others); family entropy is constant in the pooled run-level diagnostic. Mardia skewness has $p=5.81\times10^{-92}$ and kurtosis has $p=5.09\times10^{-12}$. These diagnostics validate our decision to treat single-Gaussian OOD/LLR as interpretable baselines rather than literal traffic laws. The conditional mixtures and kernel baseline address some heterogeneity, although the kernel method's calibrated threshold is too conservative to produce attacks alarms in this study.

\subsection{Service Behavior and Detector Overhead}
\begin{figure}[t]
\centering
\includegraphics[width=\columnwidth]{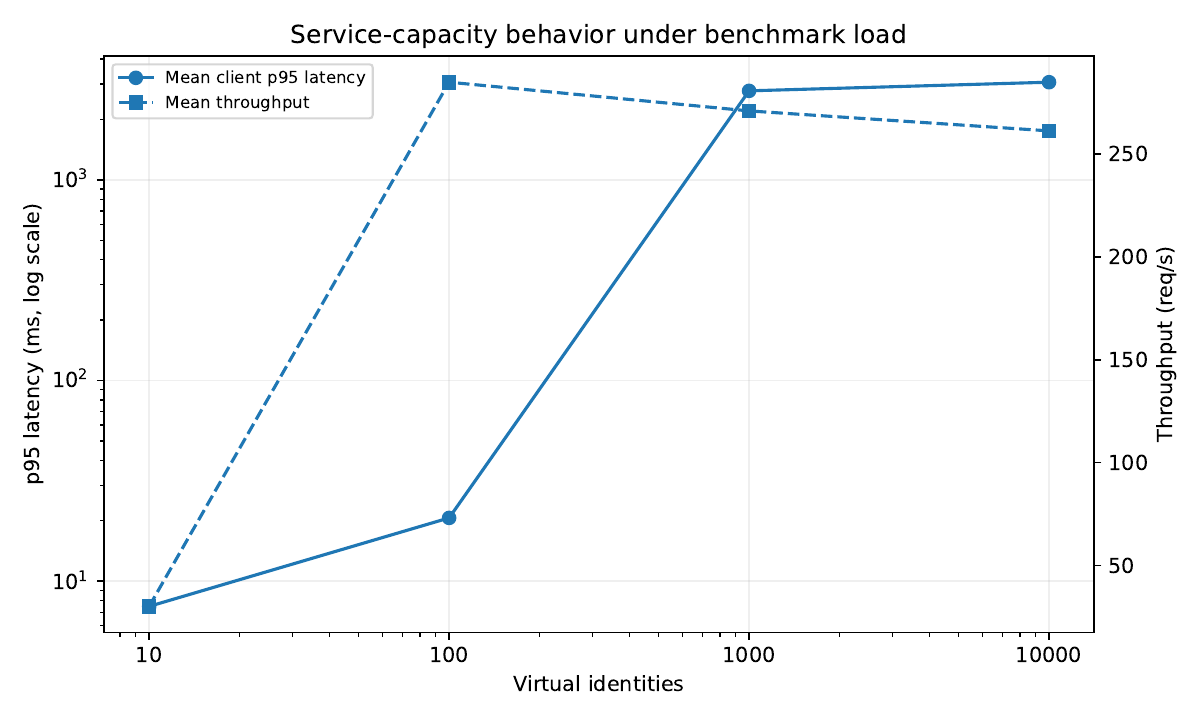}
\caption{Measured service p95 latency and throughput versus population size. The synthetic service saturates at high scale; security conclusions therefore compare attacks with matched benign load rather than claiming a production SLA.}
\label{fig:service}
\end{figure}

Figure~\ref{fig:service} separates semantic reconnaissance from volumetric denial of service. At 10 and 100 identities, mean p95 latency is approximately 7.5 and 20.7 ms. At 1,000 and 10,000 identities, the single synthetic service saturates and mean p95 latency increases to approximately 2.78 and 3.06 s, with mean throughput near 271 and 261 requests/s. Importantly, the attack-vs-matched-benign mean p95 delta at 10,000 identities is only about 1.3\%, indicating that the attack regimes do not require substantial incremental service degradation once load is matched. This supports the conceptual distinction from DDoS, but it does \emph{not} justify claiming that the 10,000-identity service is itself within a production SLA; it is capacity-limited in this laboratory configuration.

The detector's average processing cost decreases from roughly 178 $\mu$s/event at the smallest scale to 61 $\mu$s/event at 10,000 identities because fixed per-window overhead is amortized. Mean p95 processing lag remains around 5--7 ms. No test run has a transport error after bounded retry recovery, and all telemetry integrity checks pass.

\section{Discussion}
\label{sec:discussion}

\subsection{What the Results Establish}
The strongest evidence is that collective semantic behavior can be distinguishable even when the experimental per-identity budget rule never needs to be the primary signal. The two fitted reconnaissance policies are almost perfectly separable from four diverse benign controls. This confirms the core hypothesis that population-level semantic organization can expose behavior that request volume alone does not explain.

The larger contribution, however, is the failure mode. A detector trained on independent and coordinated reconnaissance does not automatically generalize to diagnostic-free, marginal-matched, camouflaged, low-and-slow, or mixed-membership strategies. Gaussian LLR is the clearest example: 100\% known-policy detection at zero observed FPR falls to 3\% unseen-policy detection. This gap is not an implementation inconvenience; it is a realistic security property. An adaptive adversary can change how a goal is achieved while preserving the goal itself.

Sequential accumulation is the most promising direction in the present results. CUSUM achieves the best overall recall and F1, and hybrid CUSUM substantially improves at 10,000 identities. This supports the idea that DCR may be better modeled as weak evidence accumulated over time than as a single anomalous window. Future work should calibrate sequential detectors under variable stream horizons and use explicit time-to-harm objectives.

\subsection{Exposure Matters More Than AUROC Alone}
A detector with high AUROC can still be operationally weak if the alarm occurs after substantial semantic acquisition. Conversely, a conservative detector may miss many attacks but alarm very early on those it detects. Reporting EBD only for detected attacks would favor such methods. Exposure-at-alarm-or-end corrects this selection effect by assigning final exposure to misses. Future evaluation should go further and define service-specific harm thresholds $C_h$ or knowledge predicates, then measure whether $t_{alarm}<t_{harm}$.

\subsection{Implications for API Defense}
The observability limit in Section~\ref{sec:threat} implies that no traffic classifier can solve DCR when adversarial and benign observable distributions are identical. A complete architecture should therefore combine detection with at least one mechanism that changes information available to either side: contextual authorization, task-scoped capabilities, selective challenges, response minimization, uncertainty injection where semantically acceptable, or service-level exposure budgets. Zero-trust and session-aware access-control research provides complementary mechanisms \cite{syed2022zerotrust,chang2026session}. SwarmReconGuard is designed to quantify the residual observable signal and the cost of relying on behavioral detection alone.

\subsection{Agentic AI Relevance Without Overclaiming}
The motivating threat includes LLM and planner agents capable of adaptive reconnaissance, but this benchmark intentionally uses scripted policies for controlled population-scale experiments. This distinction matters. The results demonstrate a property of distributed behavioral policies, not a property unique to LLM reasoning. Agentic systems make the threat more plausible because they can plan, delegate, share state, and adapt \cite{chhabra2026agentic,addis2026llmagents}; validating transfer from scripted swarms to real constrained LLM agents is future work.

\section{Threats to Validity and Limitations}
\label{sec:limitations}

\textbf{External validity:} one synthetic municipal-style API cannot establish general performance across production services. A second independently implemented service is the highest-priority extension. Cross-service calibration and evaluation should precede deployment claims.

\textbf{Policy realism:} attack and benign behaviors are controlled scripted policies. They create difficult controls but do not reproduce all properties of real users, bots, or autonomous agents. The matched-swarm construction approximates marginal matching rather than proving equality of complete individual distributions.

\textbf{Boundary segmentation:} the lab supplies trial boundaries. Production detection would require continuous segmentation, decay, sessionization, or a rolling-horizon alarm policy. Calibration of maxima also assumes comparable monitoring horizons.

\textbf{Calibration:} the 299-stream rank procedure targets a stream-level false-alarm probability under exchangeability; workload, scale, or seasonal shifts can violate this assumption. Observed FPR and Wilson intervals are therefore reported and should supersede nominal claims.

\textbf{Synthetic exposure:} semantic coverage counts unique synthetic resource keys; it is not an information-theoretic estimate of actual secrets learned by a real adversary. The three vulnerability-discovery predicates are harmless benchmark markers.

\textbf{Service capacity:} the target saturates at 1,000--10,000 identities on the experiment host. We compare matched benign and attack load to avoid conflating DCR with DDoS, but a separate capacity study is required to make strict SLA-preservation claims.

\textbf{Graph method:} handcrafted graph evidence does not improve the current hybrid and should be regarded as a negative ablation result rather than a validated component.

\textbf{No active mitigation:} detectors are observational. They do not throttle, redact, challenge, or alter service responses. Therefore the study measures detection/exposure, not prevention benefit or user-friction cost.

\section{Reproducibility, Safety, and Artifact Availability}
\label{sec:repro}

SwarmReconGuard is available at \url{https://github.com/vtavakkoli/swarm-recon-guard}. The repository provides Docker Compose profiles, source code, deterministic scenario seeds, calibration configuration, statistical reports, raw event traces, fitted detector bundles, and replay verification. The publication profile creates HTML, CSV, JSON/JSONL, compressed event traces, integrity metadata, and exact calibration artifacts. Replay re-scores saved events and verifies peak scores, alarm decisions, request counts, and exposure without reissuing HTTP requests.

The benchmark is intentionally confined to an internal Docker network. It rejects arbitrary external targets, exposes no public target port, drops Linux capabilities, mounts no Docker socket, and contains no real exploit engine, credential theft, persistence, or destructive actions. The synthetic target contains only harmless proof conditions.

\section{Future Research}
\label{sec:future}
The results suggest five concrete extensions. First, evaluate a second and third API domain and perform train-on-A/test-on-B cross-service transfer. Second, replace global Gaussian assumptions with scale/load-conditioned mixtures, robust density estimation, and calibrated conformal or sequential alternatives. Third, introduce constrained LLM/planner agents at smaller scale while retaining scripted policies for causal and population-scale controls. Fourth, formulate an adaptive attacker--defender game in which the swarm maximizes information gain while penalizing detector score, allowing explicit evaluation of evasion. Fifth, connect detection to selective disclosure controls and quantify the joint objective
\begin{equation}
\min \left(E_D+\alpha\mathrm{FPR}+\beta\Delta\mathrm{SLA}+\gamma\mathrm{Friction}\right).
\end{equation}
The research target should not be another perfect known-policy AUROC. A more meaningful goal is high unseen-policy recall at sub-1\% FPR, low exposure-at-alarm-or-end, and negligible incremental SLA impact under a service operating below its saturation knee.

\section{Conclusion}
\label{sec:conclusion}
This paper formalized Distributed Collective Reconnaissance and introduced SwarmReconGuard, a boundary-only benchmark for detecting harmful collective exploration from individually low-budget identities. The final study executed 440 measured test runs and 3.66 million test requests with exact telemetry integrity, following independent measured training and 299-stream benign calibration. Known reconnaissance policies can be detected with extremely high accuracy: Gaussian LLR achieved 100\% known-policy detection with zero observed false positives. The more important finding is that this success does not transfer automatically to unseen attack policies. Sequential evidence accumulation improves unseen-policy recall and becomes especially effective at the largest population scale, but no evaluated method solves generalization. The current graph descriptor also fails its ablation test, and the service saturates at high load, both of which are reported explicitly. These results support a cautious conclusion: collective semantic behavior is a measurable security signal, but robust DCR defense requires exposure-aware sequential detection, stronger policy-shift generalization, and complementary authorization or disclosure controls. SwarmReconGuard provides a reproducible basis for that research rather than a claim of production-ready intrusion prevention.

\section*{Data and Code Availability}
The implementation, scenario definitions, Docker configuration, and reproducibility tooling are publicly available at \url{https://github.com/vtavakkoli/swarm-recon-guard}. The complete final report bundle used for the tables and figures is included with the manuscript submission package; the experiment can be regenerated from the public repository and the publication scenario configuration.

\end{document}